# Distributed Storage Codes Meet Multiple-Access Wiretap Channels

Dimitris S. Papailiopoulos and Alexandros G. Dimakis
Department of Electrical Engineering
University of Southern California
Los Angeles, CA 90089
Email:{papailio, dimakis}@usc.edu

*Abstract*—We consider *i)* the overhead minimization of maximum-distance separable (MDS) storage codes for the repair of a single failed node and *ii)* the total secure degrees-of-freedom (S-DoF) maximization in a multiple-access compound wiretap channel. We show that the two problems are connected. Specifically, the overhead minimization for a single node failure of an *optimal* MDS code, i.e. one that can achieve the information theoretic overhead minimum, is equivalent to maximizing the S-DoF in a multiple-access compound wiretap channel. Additionally, we show that maximizing the S-DoF in a multiple-access compound wiretap channel is equivalent to minimizing the overhead of an MDS code for the repair of a departed node. An optimal MDS code maps to a full S-DoF channel and a full S-DoF channel maps to an MDS code with minimum repair overhead for one failed node. We also state a general framework for code-to-channel and channel-to-code mappings and performance bounds between the two settings. The underlying theme for all connections presented is interference alignment (IA). The connections between the two problems become apparent when we restate IA as an optimization problem. Specifically, we formulate the overhead minimization and the S-DoF maximization as rank constrained, sum-rank and max-rank minimization problems respectively. The derived connections allow us to map repair strategies of recently discovered repair codes to beamforming matrices and characterize the maximum S-DoF for the single antenna multiple-access compound wiretap channel.

## I. INTRODUCTION

A substantial volume of recent work has focused on wireless interference alignment techniques [1] – [5] for wiretap channels [6] – [11]. For these settings, the high SNR secure capacity scaling or the capacity prelog factor is the S-DoF that measure the number of secure and interference free space, time, and frequency dimensions. Confining interference and wiretapped dimensions to the minimum subspace maximizes the S-DoF and achieves the high SNR secure capacity. Perfectly fulfilling this purpose, IA serves as a means to maximize the secure and interference free signaling dimensions.

Interestingly, IA was recently used in a different framework to achieve minimum overhead repair of failed storage nodes in MDS coded distributed storage systems [12] – [20]. For the repair problem that we study, a data node of the storage array fails and a newcomer takes its place to exactly regenerate the missing contents. Appropriate repair strategies are used to mix a sufficient amount of the remaining data so that the newcomer downloads it and reconstructs the missing contents. This repair process is equivalent to solving an underdetermined system of equations in the lost data, where the undesired file pieces generate interference to a square system of equations in the lost contents. The interference has to be subtracted and the remaining system of interest needs to be full rank. The download overhead to obtain the full rank property and erase interference is proportional to the size of the lost data plus the total number of interference dimensions. Therefore, MDS storage codes and repair strategies resulting to maximum aligned interference spaces, minimize the overhead to repair a failure in the system.

In parallel to the repair problem, we study the multiple-access compound wiretap channel. A number of non-interfering users aim to communicate with one receiver of interest, while keeping their messages secret with respect to a group of non-cooperating eavesdroppers. To maximize the perfect secrecy data rate, the beamforming strategies of each user have to be designed so that the signaling dimensions observed by the eavesdropper with the sharpest eye are minimized. That way, the number of secure links between the users and the legitimate receiver at the worst wiretapping scenario are maximized. The concept of IA comes to place when designing the beamforming strategies so that the signal spaces of the users align on top of each other causing the best eavesdropper to have access to the minimum number of communication dimensions.

**Our contributions**: Motivated by interference alignment that is employed in both cases [13], [19], [20], [8], and [7], we aim to bridge the two seemingly different settings. We establish the connection by formulating IA as a rank minimization, in the same manner as [21]: *i)* Given an MDS storage code, minimizing the overhead to exactly regenerate a failed node in the storage array is equivalent to a rank constrained, sum of ranks minimization, *ii)* for a multiple-access compound wiretap channel, the S-DoF maximization can be recast to a rank constrained, maximum-rank minimization, when the user signal spaces span all available dimensions. Then, for the class of *optimal* MDS storage codes achieving the information theoretic minimum overhead bound of [12], the overhead minimization for the repair problem is equivalent to the S-DoF maximization in a class of multiple-access compound wiretap channels. Accordingly, for a class of multiple-access compound wiretap channels, maximizing the S-DoF is equivalent to minimizing the overhead of an MDS code for the repair a failed node. Additional connections come in the form of code-to-channel, channel-to-code, and repair and beamforming strategy mappings. We further establish mappings and overhead or S-DoF bounds for general classes of codes or channels.

Although we provide constructive examples that validate the connection, we need to note that there is one important difference between the two problems: distributed storage problems require rank minimization *over a finite field* whereas S-DoF maximization *over the reals*. Surprisingly, the symbol extension technique [19] is applicable in both cases, for sufficiently large finite fields. The derived connection allows us to map repair strategies of recently discovered codes in [19] and [20] to beamforming matrices and characterize the maximum S-DoF for the single antenna multiple-access compound wiretap channel, extending the works of [8] and [7]. Bagherikaram et al. [8] asymptotically achieve the outer bound of $\frac{L-1}{L}$ S-DoF for $L$ users and the single eavesdropper case, whereas Khisti [7] achieves the same bound for the $L$ antenna MISO with multiple eavesdroppers.

## II. REPAIR OF MDS STORAGE CODES

In this section, we consider the process of exactly repairing a failed node in an MDS encoded distributed storage system [14]. We show that given the storage code, the overhead minimization for the repair of a single node failure can be recast to a rank constrained sum of ranks minimization.

Let a file $\mathbf{f}$, without loss of generality be subpacketized into $M = k(n-k)\beta$ "bits" such that $\mathbf{f} \in \mathbb{F}^{k(n-k)\beta}$ and partitioned in $k$ parts $\mathbf{f} = \begin{bmatrix} \mathbf{f}_1^T \ldots \mathbf{f}_k^T \end{bmatrix}^T$, with $\mathbf{f}_i \in \mathbb{F}^{(n-k)\beta}$, where $M$ denotes the filesize in "bits" and $\beta \in \mathbb{N}^*$ the degree of subpacketization.[1] We want to store this file with rate $\frac{k}{n} \leq 1$ across $k$ systematic and $n-k$ parity storage units with storage capacity $\alpha = \frac{M}{k}$ each. Moreover, we require to retrieve the original file by accessing any $k$ of these $n$ storage nodes. These redundancy and retrieval properties are achieved by using an $(n,k)_\beta$ MDS code to encode the file across the storage nodes. The encoding is given by $\begin{bmatrix} \mathbf{I}_{k(n-k)\beta \times k(n-k)\beta} & \mathbf{A}^T \end{bmatrix} \mathbf{f}$, where $\mathbf{I}_N$ is the $N \times N$ identity matrix and we use

$$\mathbf{A} \triangleq \begin{bmatrix} \mathbf{A}_1^{(1)} & \ldots & \mathbf{A}_1^{(n-k)} \\ \mathbf{A}_2^{(1)} & \ldots & \mathbf{A}_2^{(n-k)} \\ \vdots & \vdots & \vdots \\ \mathbf{A}_k^{(1)} & \ldots & \mathbf{A}_k^{(n-k)} \end{bmatrix} \in \mathbb{F}^{k(n-k)\beta \times (n-k)^2\beta} \quad (1)$$

to denote the given systematic $(n,k)_\beta$ MDS code, where $\mathbf{A}_i^{(p)} \in \mathbb{F}^{(n-k)\beta \times (n-k)\beta}$, for all $i \in \{1, \ldots, k\}$ and $p \in \{1, \ldots, n-k\}$. Posterior to the data encoding process $k$ systematic storage nodes $\{1, \ldots, k\}$, individually store one of the $k$ parts of the file, $\mathbf{f}_1, \ldots, \mathbf{f}_k$, respectively. Each of the $n-k$ parity nodes $\{1, \ldots, n-k\}$, stores a unique linear combination of the same $k$ file pieces. The structure of the storage array is given in Fig. 1, where each node expends exactly $\alpha = (n-k)\beta$ worth of storage capacity. Observe that $\mathbf{A}_i^{(p)} \in \mathbb{F}^{(n-k)\beta \times (n-k)\beta}$ represents a matrix of coding coefficients used by the $p$th parity node to "mix" the contents of the $i$th file piece $\mathbf{f}_i$, $i \in \{1, \ldots, k\}$, $p \in \{1, \ldots, n-k\}$.

[1]For distributed storage problems $\mathbb{F}$ corresponds to a finite field; for the sake of generality we note that $\mathbb{F}$ may represent the real numbers.

| systematic node | file data |
|---|---|
| 1 | $\mathbf{f}_1$ |
| $\vdots$ | $\vdots$ |
| $k$ | $\mathbf{f}_k$ |
| parity node | parity data |
| 1 | $\left(\mathbf{A}_1^{(1)}\right)^T \mathbf{f}_1 + \ldots + \left(\mathbf{A}_k^{(1)}\right)^T \mathbf{f}_k$ |
| $\vdots$ | $\vdots$ |
| $n-k$ | $\left(\mathbf{A}_1^{(n-k)}\right)^T \mathbf{f}_1 + \ldots + \left(\mathbf{A}_k^{(n-k)}\right)^T \mathbf{f}_k$ |

Fig. 1. STORAGE CODE

To maintain the same redundancy when a single systematic node fails, a repair process takes place to regenerate the lost data in a *newcomer* storage component. This process is carried out as linear operations on the contents of the $n-1$ remaining nodes. Let for example, an $(n,k)_\beta$ MDS code $\mathbf{A}$ and a systematic node $i \in \{1, \ldots, k\}$ of the array of $n$ nodes fail. Then, a newcomer joins the storage network and we assume it connects to all $n-1$ remaining nodes that are required to transmit to it sufficient data to reconstruct $\mathbf{f}_i$. The repair of systematic node $i \in \{1, \ldots, k\}$ can be seen as a two parallel part process. First, observe that the missing piece $\mathbf{f}_i$ exists as a term of linear combination *only* at each parity node, as Fig. 1 shows. Carrying out one part of the repair, each parity node $p \in \{1, \ldots, n-k\}$ sends a size of $\beta$ data ($\beta$ equations) to the newcomer

$$\mathbf{y}_i^{(p)} = \left(\mathbf{R}_i^{(p)}\right)^T \left(\left(\mathbf{A}_1^{(p)}\right)^T \mathbf{f}_1 + \ldots + \left(\mathbf{A}_k^{(p)}\right)^T \mathbf{f}_k\right)$$
$$= \begin{bmatrix} \mathbf{A}_1^{(p)} \mathbf{R}_i^{(p)} \ldots \mathbf{A}_k^{(p)} \mathbf{R}_i^{(p)} \end{bmatrix}^T \mathbf{f} \in \mathbb{F}^\beta,$$

where $\mathbf{R}_i^{(p)} \in \mathbb{F}^{(n-k)\beta \times \beta}$ is the *repair matrix* that is free to design and mixes the contents of parity node $p$ for the repair process of systematic node $i$, $i \in \{1, \ldots, k\}$. In the same manner, all parity nodes proceed in sending a total of $(n-k)\beta$ linear equations to the newcomer. Eventually, it receives the following stack of equations

$$\mathbf{y}_i \triangleq \begin{bmatrix} \left(\mathbf{y}_i^{(1)}\right)^T \ldots \left(\mathbf{y}_i^{(n-k)}\right)^T \end{bmatrix}$$
$$= \begin{bmatrix} \left(\mathbf{A}_1^{(1)} \mathbf{R}_i^{(1)}\right)^T & \ldots & \left(\mathbf{A}_k^{(1)} \mathbf{R}_i^{(1)}\right)^T \\ \vdots & \vdots & \vdots \\ \left(\mathbf{A}_1^{(n-k)} \mathbf{R}_i^{(n-k)}\right)^T & \ldots & \left(\mathbf{A}_k^{(n-k)} \mathbf{R}_i^{(n-k)}\right)^T \end{bmatrix} \mathbf{f} \quad (2)$$
$$= \underbrace{\begin{bmatrix} \left(\mathbf{A}_i^{(1)} \mathbf{R}_i^{(1)}\right)^T \\ \vdots \\ \left(\mathbf{A}_i^{(n-k)} \mathbf{R}_i^{(n-k)}\right)^T \end{bmatrix} \mathbf{f}_i}_{\text{useful data}} + \sum_{u=1, u \neq i}^{k} \underbrace{\begin{bmatrix} \left(\mathbf{A}_u^{(1)} \mathbf{R}_i^{(1)}\right)^T \\ \vdots \\ \left(\mathbf{A}_u^{(n-k)} \mathbf{R}_i^{(n-k)}\right)^T \end{bmatrix} \mathbf{f}_u}_{\text{interference by } \mathbf{f}_u},$$

where $\mathbf{y}_i \in \mathbb{F}^{(n-k)\beta}$. Retrieving the lost piece solely from (2) is equivalent to solving an underdetermined set of $(n-k)\beta$ equations in the $k(n-k)\beta$ unknown variables of $\mathbf{f}$, with respect to only the $(n-k)\beta$ unknowns of $\mathbf{f}_i$. Obtaining $\mathbf{f}_i$ is not possible due to the *interference* components in the received equations created by the undesired $(k-1)(n-k)\beta$ unknowns $\mathbf{f}_u$, $u \in \{1,\ldots,k\}\setminus i$, as noted in (2).

Hence, for the other part of the repair process, the newcomer needs to "erase" all interference caused by the undesired $(k-1)(n-k)\beta$ unknowns. This is possible through downloading data from the remaining systematic nodes $\{1,\ldots,k\}\setminus i$ and appropriately combining them to construct exactly the sum of interference components and regenerate $\mathbf{f}_i$ from $\mathbf{y}_i$ through the following linear operations

$$\mathbf{f}_i = \begin{bmatrix} \left(\mathbf{A}_i^{(1)}\mathbf{R}_i^{(1)}\right)^T \\ \vdots \\ \left(\mathbf{A}_i^{(n-k)}\mathbf{R}_i^{(n-k)}\right)^T \end{bmatrix}^{-1} \left( \mathbf{y}_i - \sum_{u=1,u\neq i}^{k} \begin{bmatrix} \left(\mathbf{A}_u^{(1)}\mathbf{R}_i^{(1)}\right)^T \\ \vdots \\ \left(\mathbf{A}_u^{(n-k)}\mathbf{R}_i^{(n-k)}\right)^T \end{bmatrix} \mathbf{f}_u \right). \tag{3}$$

To uniquely determine $\mathbf{f}_i$, it is required that

$$\mathrm{rank}\left(\begin{bmatrix} \mathbf{A}_i^{(1)}\mathbf{R}_i^{(1)} \ldots \mathbf{A}_i^{(n-k)}\mathbf{R}_i^{(n-k)} \end{bmatrix}\right) = (n-k)\beta, \tag{4}$$

therefore it is necessary that all downloaded equations of (2) are linearly independent. Surprisingly, the size of data needed to be downloaded from the systematic nodes to erase interference is not necessarily equal to the size of interfering components $(k-1)\cdot(n-k)\beta$, but depends on the level of alignment of such interference. Namely, to erase interference created by $\mathbf{f}_u$, we need only download from the systematic part $u$ the smallest number of equations that can generate the interference components: a size equal the dimension of a linearly independent basis of the row span of the interference space $\begin{bmatrix} \mathbf{A}_u^{(1)}\mathbf{R}_i^{(1)} \ldots \mathbf{A}_u^{(n-k)}\mathbf{R}_i^{(n-k)} \end{bmatrix}$, $u \in \{1,\ldots,k\}\setminus i$, suffices.

Consequently, the size of data needed to be downloaded from the systematic parts to erase interference, equals exactly the sum of ranks of the interference spaces

$$\sum_{u=1,u\neq i}^{k} \mathrm{rank}\left(\begin{bmatrix} \mathbf{A}_u^{(1)}\mathbf{R}_i^{(1)} \ldots \mathbf{A}_u^{(n-k)}\mathbf{R}_i^{(n-k)} \end{bmatrix}\right).$$

Therefore, for an $(n,k)_\beta$ MDS code $\mathbf{A}$ and a certain set of repair matrices

$$\mathbf{R}_i \triangleq \begin{bmatrix} \mathbf{R}_i^{(1)} \ldots \mathbf{R}_i^{(n-k)} \end{bmatrix} \in \mathbb{F}^{(n-k)\beta \times (n-k)^2\beta} \tag{5}$$

the overhead for exact repair of a systematic node $i \in \{1,\ldots,k\}$ is defined as

$$\delta\left((n,k)_\beta, \mathbf{P}_i\mathbf{A}, \mathbf{R}_i\right)$$
$$\triangleq \frac{(n-k)\beta + \sum_{u=1,u\neq i}^{k} \mathrm{rank}\left(\begin{bmatrix} \mathbf{A}_u^{(1)}\mathbf{R}_i^{(1)} \ldots \mathbf{A}_u^{(n-k)}\mathbf{R}_i^{(n-k)} \end{bmatrix}\right)}{(n-k)\beta}$$
$$= 1 + \sum_{u=1,u\neq i}^{k} \frac{\mathrm{rank}\left(\begin{bmatrix} \mathbf{A}_u^{(1)}\mathbf{R}_i^{(1)} \ldots \mathbf{A}_u^{(n-k)}\mathbf{R}_i^{(n-k)} \end{bmatrix}\right)}{(n-k)\beta}, \tag{6}$$

where $\mathbf{P}_i \triangleq \begin{bmatrix} \mathbf{E}_i^T \mathbf{E}_1^T \ldots \mathbf{E}_{i-1}^T \mathbf{E}_{i+1}^T \ldots \mathbf{E}_k^T \end{bmatrix}^T$, $\mathbf{E}_i \triangleq \mathbf{e}_i \otimes \mathbf{I}_{(n-k)\beta}$, and $\mathbf{e}_i$ corresponds to the $i$th column of $\mathbf{I}_k$, for all $i \in \{2,\ldots,k\}$. Observe that for an $(n,k)_\beta$ MDS code $\mathbf{A}$ we have

$$\mathbf{P}_i\mathbf{A} = \begin{bmatrix} \mathbf{A}_i^{(1)} & \ldots & \mathbf{A}_i^{(n-k)} \\ \hline \mathbf{A}_1^{(1)} & \ldots & \mathbf{A}_1^{(n-k)} \\ \vdots & \vdots & \vdots \\ \mathbf{A}_{i-1}^{(1)} & \ldots & \mathbf{A}_{i-1}^{(n-k)} \\ \mathbf{A}_{i+1}^{(1)} & \ldots & \mathbf{A}_{i+1}^{(n-k)} \\ \vdots & \vdots & \vdots \\ \mathbf{A}_k^{(1)} & \ldots & \mathbf{A}_k^{(n-k)} \end{bmatrix},$$

i.e. the first row block corresponds to the set of coding matrices multiplying file piece $\mathbf{f}_i$, $i \in \{1,\ldots,k\}$, for $i \in \{1,\ldots,k\}$.

Therefore, the overhead is the ratio of the number of downloaded equations to the size of the lost piece. This ratio is minimized when the *interference alignment* is maximized, i.e. when the row spans of the matrices generating the interference belong to a subspace of minimum dimension. To minimize the overhead $\delta((n,k)_\beta, \mathbf{P}_i\mathbf{A}, \mathbf{R}_i)$, for a given $(n,k)_\beta$ MDS storage code $\mathbf{A}$, we can only optimize over the repair matrices $\mathbf{R}_i$. This optimization can be recast to the following rank constrained, sum of ranks minimization that has to be performed over $\mathbb{F}$

$$\mathcal{R}((n,k)_\beta, \mathbf{P}_i\mathbf{A}):$$
$$\min_{\mathbf{R}_i^{(1)},\ldots,\mathbf{R}_i^{(n-k)}} \sum_{u=1,u\neq i}^{k} \mathrm{rank}\left(\begin{bmatrix} \mathbf{A}_u^{(1)}\mathbf{R}_i^{(1)} \ldots \mathbf{A}_u^{(n-k)}\mathbf{R}_i^{(n-k)} \end{bmatrix}\right)$$
$$\text{s.t.: } \mathrm{rank}\left(\begin{bmatrix} \mathbf{A}_i^{(1)}\mathbf{R}_i^{(1)} \ldots \mathbf{A}_i^{(n-k)}\mathbf{R}_i^{(n-k)} \end{bmatrix}\right) = (n-k)\beta.$$

*Lemma 1:* For a given $(n,k)_\beta$ MDS storage code $\mathbf{A}$, solving $\mathcal{R}((n,k)_\beta, \mathbf{P}_i\mathbf{A})$ is equivalent to minimizing the overhead to repair node $i \in \{1,\ldots,k\}$.

We note that the minimum repair overhead for a code of a given rate drastically depends on its design. The cut-set analysis of [12] defines *optimal* $(n,k)_\beta$ MDS storage codes that achieve the information theoretic minimum repair overhead: each of the $n-1$ surviving nodes has to deliver a size of exactly $\beta$ repair data. This means that when repair has to be performed for these storage codes, we can always find repair matrices such that interference spaces align in $\beta$-dimensional subspaces, resulting to minimum repair overhead $\frac{n-1}{n-k}$.

We now shift to a seemingly unrelated problem: the S-DoF maximization of a multiple-access compound wiretap channel. Then, we establish a formal connection to the repair problem.

III. MULTIPLE-ACCESS COMPOUND WIRETAP CHANNEL

In this section, we consider a multiple-access compound wiretap channel. We show that given full rank signal spaces, the S-DoF maximization can be formulated as a rank constrained maximum rank minimization.

In the multiple-access compound wiretap channel, $L$ users wish to communicate with a sole legitimate receiver, while maintaining message secrecy with respect to $K-1$ eavesdroppers. Each user $l \in \{1,\ldots L\}$ is equipped with $M_\mathrm{t}$ transmit antennas and wishes to transmit a length $N$ symbol vector

$\mathbf{x}^{(l)} \in \mathbb{R}^{N \times 1}$. We assume that the legitimate receiver and each eavesdropper $v$ is equipped with $M_r$ receive antennas, respectively, where $v \in \{1, \ldots, K-1\}$. We further assume that $M_t = M_r = LN$. We denote such channel as an $(L,N)^{K-1}$ multiple-access compound wiretap channel. The $LN \times 1$ pulse matched and downconverted received signals at the legitimate receiver and the $v$th eavesdropper are

$$\mathbf{y} = \sum_{l=1}^{L} \mathbf{H}^{(l)} \mathbf{V}^{(l)} \mathbf{x}^{(l)} + \mathbf{w} \quad (7)$$

and $\mathbf{y}_{ev} = \sum_{l=1}^{L} \mathbf{H}_{ev}^{(l)} \mathbf{V}^{(l)} \mathbf{x}^{(l)} + \mathbf{w}_{ev}, \quad (8)$

respectively, where $\mathbf{H}^{(l)} \in \mathbb{R}^{LN \times LN}$ represents the "channel processing" between the $l$th user and the receiver of interest, $\mathbf{V}^{(l)} \in \mathbb{R}^{LN \times N}$ is the beamforming matrix of user $l$, and $\mathbf{w}$ accounts for zero-mean additive white Gaussian noise with covariance matrix $\sigma^2 \mathbf{I}_{LN}$, where $v \in \{1, \ldots, K-1\}$ and $l \in \{1, \ldots, L\}$. $\mathbf{H}_{ev}^{(l)} \in \mathbb{R}^{LN \times LN}$ represents the channel between the $l$th user and eavesdropper $v$ and $\mathbf{w}_{ev}$ corresponds to the zero-mean additive white Gaussian noise with covariance matrix $\sigma^2 \mathbf{I}_{LN}$, for $v \in \{1, \ldots, K-1\}$ and $l \in \{1, \ldots, L\}$. We further assume a power constraint on the transmitted signal of each user $E\{\|\mathbf{V}^{(l)}\mathbf{x}^{(l)}\|_2^2\} \leq P$ and $\left(\mathbf{V}^{(l)}\right)^T \mathbf{V}^{(l)} = \frac{P}{N}\mathbf{I}_N$, for some $P > 0$, and all $l \in \{1, \ldots, L\}$. We assume that the elements of all channel matrices are drawn i.i.d. from a continuous distribution and at each channel use all receivers and transmitters have perfect channel knowledge.

We proceed by rewriting (7) and (8) as

$$\mathbf{y} = \begin{bmatrix} \mathbf{H}^{(1)}\mathbf{V}^{(1)} \ldots \mathbf{H}^{(L)}\mathbf{V}^{(L)} \end{bmatrix} \mathbf{x} + \mathbf{w}$$

and $\mathbf{y}_{ev} = \begin{bmatrix} \mathbf{H}_{ev}^{(1)}\mathbf{V}^{(1)} \ldots \mathbf{H}_{ev}^{(L)}\mathbf{V}^{(L)} \end{bmatrix} \mathbf{x} + \mathbf{w}_{ev}$

and defining

$$\mathbf{H} \triangleq \begin{bmatrix} \mathbf{H}^{(1)} & \ldots & \mathbf{H}^{(L)} \\ \hline \mathbf{H}_{e1}^{(1)} & \ldots & \mathbf{H}_{e1}^{(L)} \\ \vdots & \vdots & \vdots \\ \mathbf{H}_{e(K-1)}^{(1)} & \ldots & \mathbf{H}_{e(K-1)}^{(L)} \end{bmatrix} \in \mathbb{R}^{KLN \times L^2 N}, \quad (9)$$

which denotes an instance of an $(L,N)^{K-1}$ multiple-access compound wiretap channel. We further define the concatenation of all $L$ beamforming matrices

$$\mathbf{V} \triangleq \begin{bmatrix} \mathbf{V}^{(1)} \ldots \mathbf{V}^{(L)} \end{bmatrix} \in \mathbb{R}^{LN \times LN}. \quad (10)$$

Then, to achieve the secure sum capacity for perfect secrecy of an $(L,N)^{K-1}$ multiple-access compound wiretap channel at the high SNR regime, we have to maximize the (normalized) total S-DoF. For this case the total S-DoF of an $(L,N)^{K-1}$ multiple-access compound wiretap channel $\mathbf{H}$, for a given set of beamforming matrices $\mathbf{V}$, and assuming $\mathbf{x}^{(l)}$, $l \in \{1, \ldots, L\}$, is a zero-mean real Gaussian vector with covariance matrix $\mathbf{I}_N$, is given by

$\eta\left((L,N)^{K-1}, \mathbf{H}, \mathbf{V}\right)$ (11)

$\triangleq \frac{\left[\text{rank}\left(\begin{bmatrix}\mathbf{H}^{(1)}\mathbf{V}^{(1)}\ldots\mathbf{H}^{(L)}\mathbf{V}^{(L)}\end{bmatrix}\right) - \max_{v \in \{1,\ldots,K-1\}} \text{rank}\left(\begin{bmatrix}\mathbf{H}_{ev}^{(1)}\mathbf{V}^{(1)}\ldots\mathbf{H}_{ev}^{(L)}\mathbf{V}^{(L)}\end{bmatrix}\right)\right]^+}{LN}$

where $[a]^+ = \begin{cases} a, & a > 0, \\ 0, & a \leq 0 \end{cases}$. Hence, the set of beamforming matrices $\mathbf{V}$ has to be designed such that the legitimate signal space spans the maximum dimensions possible and each of the $K-1$ eavesdropper signal spaces has to collapse in as small dimensions as possible, such that the maximum number of dimensions among the eavesdropper spaces is minimized.

Observe that given a signal space $\begin{bmatrix}\mathbf{H}^{(1)}\mathbf{V}^{(1)}\ldots\mathbf{H}^{(L)}\mathbf{V}^{(L)}\end{bmatrix}$ spanning $LN$ dimensions, the S-DoF of an $(L,N)^{K-1}$ multiple-access compound wiretap channel $\mathbf{H}$ can be maximized by solving the following optimization problem

$$\boxed{\begin{aligned}&\mathcal{V}\left((L,N)^{K-1}, \mathbf{H}\right): \\ &\min_{\mathbf{V}^{(1)},\ldots,\mathbf{V}^{(L)}} \max_{v \in \{1,\ldots,K-1\}} \text{rank}\left(\begin{bmatrix}\mathbf{H}_{ev}^{(1)}\mathbf{V}^{(1)}\ldots\mathbf{H}_{ev}^{(L)}\mathbf{V}^{(L)}\end{bmatrix}\right) \\ &\text{s.t.: rank}\left(\begin{bmatrix}\mathbf{H}^{(1)}\mathbf{V}^{(1)}\ldots\mathbf{H}^{(L)}\mathbf{V}^{(L)}\end{bmatrix}\right) = LN.\end{aligned}}$$

*Lemma 2:* For a given $(L,N)^{K-1}$ multiple-access compound wiretap channel $\mathbf{H}$, solving $\mathcal{V}\left((L,N)^{K-1}, \mathbf{H}\right)$ is equivalent to maximizing the total S-DoF, when the legitimate signal space spans $LN$ dimensions.

We note that the minimum rank of an eavesdropper's observable space cannot be less than $N$ assuming that the beamforming matrices are full column rank due to each beamforming matrix being full rank. The following bound

$$\eta\left((L,N)^{K-1}, \mathbf{H}, \mathbf{V}\right) \leq \frac{LN - N}{LN} = \frac{L-1}{L} \quad (12)$$

always serves as an outerbound to the achievable total S-DoF.

In the next section, we establish a connection between the S-DoF maximization and the repair overhead minimization.

## IV. ESTABLISHING THE CONNECTIONS

The two models of sections II and III correspond to physically unrelated settings. However, we observe in the following that the two problems share the same mathematical formulation. This allows us to establish a connection between the overhead minimization and the S-DoF maximization.

We begin by observing that for both overhead minimization and S-DoF maximization setups, there exist two types of spaces with respect to the respective objectives: the "useful" spaces and the "harmful" spaces. The useful spaces need always be full rank for both problems. Then, the sum of dimensions of the set of harmful spaces for the first problem, or the maximum dimension of harmful space for the second, need to be minimized respectively. Specifically, for the overhead minimization problem, the useful space when we repair node $i$ is $\begin{bmatrix}\mathbf{A}_i^{(1)}\mathbf{R}_i^{(1)}\ldots\mathbf{A}_i^{(n-k)}\mathbf{R}_i^{(n-k)}\end{bmatrix}$, while the $k-1$ interference spaces $\begin{bmatrix}\mathbf{A}_u^{(1)}\mathbf{R}_i^{(1)}\ldots\mathbf{A}_u^{(n-k)}\mathbf{R}_i^{(n-k)}\end{bmatrix}$, $u \in \{1,\ldots,k\}\setminus i$, consist harmful components; the larger their sum of dimensions, the larger the overhead to repair systematic node $i \in \{1,\ldots,k\}$. Then, for the S-DoF maximization, the useful space is $\begin{bmatrix}\mathbf{H}^{(1)}\mathbf{V}^{(1)}\ldots\mathbf{H}^{(L)}\mathbf{V}^{(L)}\end{bmatrix}$, while there exist $K-1$ eavesdropper spaces $\begin{bmatrix}\mathbf{H}_{ev}^{(1)}\mathbf{V}^{(1)}\ldots\mathbf{H}_{ev}^{(L)}\mathbf{V}^{(L)}\end{bmatrix}$,

| Overhead Minimization for Repair of Node $i$ | | S-DoF Maximization | |
|---|---|---|---|
| Spaces | | | |
| Useful data space | $\mathbf{A}_i^{(1)}\mathbf{R}_i^{(1)}\ldots\mathbf{A}_i^{(n-k)}\mathbf{R}_i^{(n-k)}$ | $\mathbf{H}^{(1)}\mathbf{V}^{(1)}\ldots\mathbf{H}^{(L)}\mathbf{V}^{(L)}$ | Legitimate receiver space |
| Interference space | $\mathbf{A}_u^{(1)}\mathbf{R}_i^{(1)}\ldots\mathbf{A}_u^{(n-k)}\mathbf{R}_i^{(n-k)}$ | $\mathbf{H}_{ev}^{(1)}\mathbf{V}^{(1)}\ldots\mathbf{H}_{ev}^{(L)}\mathbf{V}^{(L)}$ | Eavesdropper's space |
| Matrices | | | |
| Code matrix | $\mathbf{P}_i\mathbf{A}$ | $\mathbf{H}$ | Channel matrix |
| Repair strategy | $\mathbf{R}_i$ | $\mathbf{V}$ | Beamforming strategy |
| Dimensions | | | |
| # Parity nodes | $n-k$ | $L$ | # Users |
| # Interference spaces | $k-1$ | $K-1$ | # Wiretapped spaces |
| Subpacketization | $\beta$ | $N$ | # Symbols/User |
| Constraints | | | |
| $\mathrm{rank}\left(\left[\mathbf{A}_i^{(1)}\mathbf{R}_i^{(1)}\ldots\mathbf{A}_i^{(n-k)}\mathbf{R}_i^{(n-k)}\right]\right)=(n-k)\beta$ | | $\mathrm{rank}\left(\left[\mathbf{H}^{(1)}\mathbf{V}^{(1)}\ldots\mathbf{H}^{(L)}\mathbf{V}^{(L)}\right]\right)=LN$ | |
| Cost functions$^{(*)}$ | | | |
| $\sum_{u=1,u\neq i}^{k}\mathrm{rank}\left(\left[\mathbf{A}_u^{(1)}\mathbf{R}_i^{(1)}\ldots\mathbf{A}_u^{(n-k)}\mathbf{R}_i^{(n-k)}\right]\right)$ | | $\min_{v\in\{1,\ldots,K-1\}}\mathrm{rank}\left(\left[\mathbf{H}_{ev}^{(1)}\mathbf{V}^{(1)}\ldots\mathbf{H}_{ev}^{(L)}\mathbf{V}^{(L)}\right]\right)$ | |

Fig. 2. ANALOGIES

$v \in \{1,\ldots,K-1\}$, that potentially harm the objective; the larger the maximum of their dimensions, the lower the S-DoF.

Furthermore, we observe that each parity node $p \in \{1,\ldots,n-k\}$ and each user $l \in \{1,\ldots,L\}$ is associated with a repair matrix $\mathbf{R}_i^{(p)}$ and a beamforming matrix $\mathbf{V}^{(l)}$, respectively. The repair and beamforming matrices multiply i) one matrix ($\mathbf{A}_i^{(p)}$ or $\mathbf{H}^{(l)}$ respectively) contributing to useful space and ii) $k-1$ or $K-1$ matrices ($\mathbf{A}_u^{(p)}$, $u \in \{1,\ldots,k\}\setminus i$, or $\mathbf{H}_{ev}^{(l)}$, $v \in \{1,\ldots,K-1\}$, respectively) contributing to harmful spaces. Then, given an $(n,k)_\beta$ MDS code $\mathbf{A}$, we optimize over the $n-k$ matrices $\mathbf{R}_i^{(1)},\ldots,\mathbf{R}_i^{(n-k)}$ to minimize the sum of interference space dimensions; that way the overhead to repair node $i$ is minimized. Analogously, for a given $(L,N)^{K-1}$ multiple-access compound wiretap channel $\mathbf{H}$, we optimize over $L$ matrices $\mathbf{V}^{(1)},\ldots,\mathbf{V}^{(L)}$ so that the largest eavesdropper space is minimized; hence, the S-DoF is maximized.

Interestingly, we use these mathematical formulation analogies to prove that the two problems are equivalent and direct code-to-channel or channel-to-code mappings are possible when the codes considered are optimal or the channels achieve the outerbound of (12). When considering conventional MDS codes and more general multiple-access compound wiretap channels, we establish performance bounds in terms of repair overhead and achievable S-DoF. Before we proceed, it is important note that *i)* in contrast to storage codes, channels are given by nature and *ii)* for equivalencies and mappings to hold calculations have to be performed over the same field. All analogies stated in this section are given in Fig. 2. (∗) denotes that the analogy holds for optimal MDS codes and full S-DoF channels.

### A. Equivalent Problems

In the following, we state the connections between the repair overhead minimization for optimal MDS codes with the S-DoF maximization for channels that achieve (12). The connections are articulated through theoreoms that $\mathcal{R}$ and $\mathcal{V}$ are equivalent, code-to-channel and channel-to-code mappings for MDS codes and multiple-access compound wiretap channels, and constructive examples.[2]

*Theorem 1:* Let an optimal $(n,k)_\beta$ MDS code $\mathbf{A}$. Then, $\mathcal{R}\left((n,k)_\beta,\mathbf{P}_i\mathbf{A}\right)$ is equivalent to $\mathcal{V}\left((n-k,\beta)^{k-1},\mathbf{P}_i\mathbf{A}\right)$, for all $i \in \{1,\ldots,k\}$.

*Corollary 1:* Let an optimal $(n,k)_\beta$ MDS code $\mathbf{A}$, defined over $\mathbb{R}$. Then, $\mathbf{A}$ maps to $k$, $(n-k,\beta)^{k-1}$ multiple-access compound wiretap channel instances $\mathbf{H} = \mathbf{P}_i\mathbf{A}$, $\forall i \in \{1,\ldots,k\}$, where the outer bound of $\frac{n-k-1}{n-k}$ S-DoF is achievable.

Following Theorem 1, we establish that the S-DoF maximization for the class of multiple-access compound wiretap channels where (12) is achievable, is equivalent to the repair overhead minimization of a storage code.

*Theorem 2:* Let an $(L,N)^{K-1}$ multiple-access compound wiretap channel $\mathbf{H}$, where $\frac{L-1}{L}$ total S-DoF is achievable. Then, $\mathcal{V}\left((L,N)^{K-1},\mathbf{H}\right)$ is equivalent to $\mathcal{R}\left((L+K,K)_N,\mathbf{H}\right)$.

*Corollary 2:* Let an $(L,N)^{K-1}$ multiple-access compound wiretap channel $\mathbf{H}$, where $\frac{L-1}{L}$ S-DoF is achievable. Then, $\mathbf{H}$ maps to an $(L+K,K)_N$ MDS code $\mathbf{A} = \mathbf{H}$, where the overhead to repair systematic node 1 is $\frac{n-1}{n-k}$.

*Remark*: If the elements of $\mathbf{H}$ are drawn i.i.d. from a continuous distribution, then $\mathbf{H}$ maps to an optimal $(L+K,K)_N$ MDS code $\mathbf{A} = \mathbf{H}$.

Theorems 1 and 2 explicitly state that if there existed a black-box that could solve $\mathcal{V}\left((L,N)^{K-1},\mathbf{H}\right)$, then this box could be used to solve $\mathcal{R}\left((n,k)_\beta,\mathbf{P}_i\mathbf{A}\right)$ for optimal MDS codes. A vice versa statement holds for multiple-access compound wiretap channels where (12) is achievable. We establish that optimal MDS codes map to channels where (12) is achievable. Accordingly, channels where (12) is tight map to MDS codes where the repair of a single node can be performed with minimum overhead.

---
[2]Henceforth, we assume that an algorithm solving $\mathcal{R}$ over a finite field $\mathbb{F}$ can be converted to one over $\mathbb{R}$, for any $i \in \{1,\ldots,k\}$, and an algorithm solving $\mathcal{V}$ over $\mathbb{R}$ can be converted to one over $\mathbb{F}$.

For the channel-to-code mapping any channel that satisfies the MDS properties maps to a code (over the reals) whose structure is always of practical interest. In sharp contrast, the code-to-channel mapping is of practical interest only when the structure of $\mathbf{A}$ is not artificial and can represent realistic $(n-k, \beta)^{k-1}$ multiple-access compound wiretap channel structures. The most interesting example comes for free after applying Corollary 1 on the asymptotically optimal MDS code presented in [19] and [20]. We use the code-to-channel mapping to show that the outerbound of (12) is asymptotically achievable and establish the S-DoF of the single antenna multiple-access compound wiretap channel; this is possible by mapping the repair matrices to the beamforming matrices used by the users of such system.

*Example 1*: In this example, we show that $\frac{L-1}{L}$ S-DoF is achievable for the single antenna, time-varying, multiple-access compound wiretap channel, with $L$ users and $K-1$ eavesdroppers. To achieve this result, we use symbol extension and set as beamforming matrices, the repair matrices used for the asymptotically optimal MDS code presented in [19] and [20], that maps to the same channel structure when $\mathbb{F} = \mathbb{R}$. Most interestingly, in [19] and [20] the authors prove the asymptotic achievability of the information theoretic minimum overhead for the exact repair of departed nodes for MDS codes of any rate, for $\beta \to \infty$. Although this MDS code was presented for finite fields, through this example we show that the mapping also works over the reals and this is due to the special structure of the code and repair matrices. Moving to the single antenna, time-varying, $L$ users, $K-1$ eavesdroppers, multiple-access compound wiretap channel, the received signals at the legitimate receiver and the $k$th eavesdropper at a single channel use are

$$y(t) = \sum_{l=1}^{L} h^{(l)}(t) \hat{x}^{(l)}(t) + w(t)$$

$$\text{and } y_{\text{ev}}(t) = \sum_{l=1}^{L} h_{\text{ev}}^{(l)}(t) \hat{x}^{(l)}(t) + w_{ev}(t),$$

respectively, where $h^{(l)}(t) \in \mathbb{R}$ represents the channel between the $l$th user and the legitimate receiver at time $t$, $h_{\text{ev}}^{(l)}(t) \in \mathbb{R}$ the channel between the $l$th user and $v$th eavesdropper, and $w(t)$ and $w_{ek}(t)$ account for zero-mean additive white Gaussian noise with variance $\sigma^2$ at the legitimate receiver and eavesdropper $v$, where $l \in \{1, \ldots, L\}$ and $v \in \{1, \ldots, K-1\}$. $\hat{x}^{(l)}(t) \in \mathbb{R}$ is the sumbol that user $l \in \{1, \ldots, L\}$ transmits.

Then, we employ $LN = L\Delta^{(K-1)L}$ symbol extensions such that

$$\begin{bmatrix} \hat{x}^{(l)}(1) \\ \vdots \\ \hat{x}^{(l)}(LN) \end{bmatrix} = \mathbf{V}^{(l)}\mathbf{x}^{(l)}, \mathbf{H}^{(l)} = \begin{bmatrix} h^{(l)}(1) \ldots & 0 \\ \vdots & \ddots & \vdots \\ 0 & \ldots h^{(l)}(LN) \end{bmatrix}, \text{ and}$$

$$\mathbf{H}_{\text{ev}}^{(l)} = \begin{bmatrix} h_{\text{ev}}^{(l)}(1) \ldots & 0 \\ \vdots & \ddots & \vdots \\ 0 & \ldots h_{\text{ev}}^{(l)}(LN) \end{bmatrix},$$

for $\Delta \in \mathbb{N}^+$, $l \in \{1, \ldots, L\}$, and $v \in \{1, \ldots, K-1\}$. Observe that we may perceive this symbol extended channel as an $(L, \Delta^{(K-1)L})^{K-1}$ multiple-access compound wiretap channel, where the channel matrices are diagonal. The structure of the individual channel matrices is in accordance with the diagonal structure of the coding matrices in [19] and [20]. Specifically, every code $(n, k)_\beta$ $\mathbf{A}$ of [19] and [20], with diagonal elements drawn from a continuous distribution, maps to an $(L, N)^{K-1}$ channel with matrices having the same diagonal structure described earlier, for $n - k = L$, $\beta = N$, and $k = K$. For our setting we use as beamforming matrices the repair matrices of [19] and [20], that achieve the minimum repair overhead. Namely,

$$\hat{\mathbf{V}} = \mathbf{V}^{(l)} = \left\{ \left( \prod_{l'=1}^{L} \prod_{v=1}^{K-1} \left( \mathbf{H}_{\text{ev}}^{(l')} \right)^{\alpha_{l',v}} \right) \mathbf{w} : \alpha_{l',v} \in \{1, \ldots, \Delta\} \right\},$$

for all $l \in \{1, \ldots, L\}$, where we assume the elements of $\mathbf{w} \in \mathbb{R}^{L\Delta^{(K-1)L}}$ to be drawn i.i.d. from a continuous distribution. Hence, $\hat{\mathbf{V}}$ has $\Delta^{(K-1)L}$ linearly independent column vectors almost surely. Then, we obtain the following

$$\text{rank}\left( \left[ \mathbf{H}^{(1)}\mathbf{V} \ldots \mathbf{H}^{(L)}\mathbf{V} \right] \right) = L\Delta^{(K-1)L}$$

and $\Delta^{(K-1)L} < \text{rank}\left( \left[ \mathbf{H}_{\text{ev}}^{(1)}\mathbf{V} \ldots \mathbf{H}_{\text{ev}}^{(L)}\mathbf{V} \right] \right) < (\Delta+1)^{(K-1)L}$,

with probability 1 for any $v \in \{1, \ldots, K-1\}$, when the channel coefficients are drawn i.i.d. from a continuous distribution. For this channel we achieve the following S-DoF

$$\eta\left( \left( L, \Delta^{(K-1)L} \right)^{K-1}, \mathbf{H}, \mathbf{V} \right) = \frac{L\Delta^{(K-1)L} - (\Delta+1)^{(K-1)L}}{L\Delta^{(K-1)L}}$$

$$\xrightarrow{\Delta \to \infty} \frac{L-1}{L} \quad (13)$$

almost surely and asymptotically match the outer bound of (12) for $\Delta \to \infty$. Interestingly, this result is not a function of the number of eavesdroppers, which is in accordance to the MISO compound wiretap setup of [7]. As a sidenote, we observe that the single antenna case of the channel model we considered can be seen as a MISO compound wiretap system where the beamforming is done independently at each antenna element over time, still yielding the same S-DoF as in [7].

We continue with another code-to-channel mapping example. This case is a straightforward extension of the $(L, 1)^1$ multiple-access wiretap channel of [8], for users wishing to transmit more than 1 symbols.

*Example 2*: Let an $(n, 2)_\beta$ MDS code $\mathbf{A}$, where each element of $\mathbf{A}$ is drawn i.i.d. from a continuous distribution. Then, an optimal set of repair vectors for the repair of node $i$ are $\mathbf{R}_i = \left[ \left( \mathbf{A}_u^{(1)} \right)^{-1} \mathbf{W} \ldots \left( \mathbf{A}_u^{(n-k)} \right)^{-1} \mathbf{W} \right]$, where $\mathbf{W} \in \mathbb{R}^{(n-k)\beta \times \beta}$ is also drawn i.i.d. from a continuous distribution and $(i, u) \in \{(1,2), (2,1)\}$. These repair matrices yield an interference space of dimension $\beta$. Namely, for the repair of systematic node $i \in \{1, 2\}$ we have

$$\text{rank}\left( \begin{bmatrix} \left( \mathbf{A}_u^{(1)} \mathbf{R}_i^{(1)} \right)^T \\ \vdots \\ \left( \mathbf{A}_u^{(n-k)} \mathbf{R}_i^{(n-k)} \right)^T \end{bmatrix} \right) = \text{rank}\left( \begin{bmatrix} (\mathbf{W})^T \\ \vdots \\ (\mathbf{W})^T \end{bmatrix} \right) = \beta,$$

for $(i, u) \in \{(1,2), (2,1)\}$, while obeying the full rank constraint with probability 1. The storage code $\mathbf{A}$ maps to an $(n-k, \beta)^1$ multiple-access wiretap channel $\mathbf{H} = \mathbf{A}$. To achieve (12) for $\mathbf{H}$, we use as beamforming matrices the repair matrices $\hat{\mathbf{R}}_1$, i.e. $\mathbf{V}^{(l)} = \left(\mathbf{H}_{e1}^{(l)}\right)^{-1}\mathbf{W}$. Then,

$$\text{rank}\left(\begin{bmatrix}\left(\mathbf{H}_{e1}^{(1)}\mathbf{V}^{(1)}\right)^T \\ \vdots \\ \left(\mathbf{H}_{e1}^{(n-k)}\mathbf{V}^{(n-k)}\right)^T\end{bmatrix}\right) = \text{rank}\left(\begin{bmatrix}(\mathbf{W})^T \\ \vdots \\ (\mathbf{W})^T\end{bmatrix}\right) = \beta,$$

and $\text{rank}\left(\begin{bmatrix}\mathbf{H}^{(1)}\mathbf{V}^{(1)}\ldots\mathbf{H}^{(n-k)}\mathbf{V}^{(n-k)}\end{bmatrix}\right) = (n-k)\beta$, almost surely. Thus, $\frac{n-k-1}{n-k}$ S-DoF is almost surely achievable for all, but a measure zero set of, $(n-k,\beta)^1$ multiple-access wiretap channels. Since $n-k,\beta \in \mathbb{N}^+$, this result can be equivalently stated for $(L,N)^1$ channels, for any $L,N \in \mathbb{N}^+$.

Next, we exhibit a channel-to-code mapping, where we rederive the optimal $(n,2)$ MDS code of [13] using the channel matrices of [8].

*Example 3*: Let an $(L,1)^1$ multiple-access compound wiretap channel $\mathbf{H}$. Then, by setting $\mathbf{v}^{(l)} = \left(\mathbf{H}_e^{(l)}\right)^{-1}\mathbf{w}$, for $l \in \{1,\ldots,L\}$, such that $\mathbf{H}_e^{(l)}\mathbf{v}^{(l)} = \mathbf{w}$, we obtain $\begin{bmatrix}\mathbf{H}_{e1}^{(1)}\mathbf{v}^{(1)}\ldots\mathbf{H}_{e1}^{(L)}\mathbf{v}^{(L)}\end{bmatrix} = [\mathbf{w}\ldots\mathbf{w}]$, and achieve (12), almost surely for vector $\mathbf{w}$ drawn from a continuous distribution. Any such randomly drawn $(L,1)^1$ channel maps to an optimal $(L+2,2)_1$ MDS code $\mathbf{A} = \mathbf{H}$, or equivalently, to an optimal $(n,2)_1$ MDS code $\mathbf{A}$. Then, we use the same structure used before for the beamforming matrices, i.e., $\mathbf{R}_1 = \begin{bmatrix}\left(\mathbf{A}_2^{(1)}\right)^{-1}\mathbf{w}\ldots\left(\mathbf{A}_2^{(L)}\right)^{-1}\mathbf{w}\end{bmatrix}$ and $\mathbf{R}_2 = \begin{bmatrix}\left(\mathbf{A}_1^{(1)}\right)^{-1}\mathbf{w}\ldots\left(\mathbf{A}_1^{(L)}\right)^{-1}\mathbf{w}\end{bmatrix}$. Observe that these repair matrices yield an interference space of dimension $\beta = 1$ for the repair of systematic node $i \in \{1,2\}$, that is

$$\text{rank}\left(\begin{bmatrix}\left(\mathbf{A}_u^{(1)}\mathbf{R}_i^{(1)}\right)^T \\ \vdots \\ \left(\mathbf{A}_u^{(n-k)}\mathbf{R}_i^{(n-k)}\right)^T\end{bmatrix}\right) = \text{rank}\left(\begin{bmatrix}(\mathbf{w})^T \\ \vdots \\ (\mathbf{w})^T\end{bmatrix}\right) = 1,$$

and the information theoretic minimum repair overhead $\frac{L+1}{L}$ is achieved.

### B. General Performance Bounds

Here, we extend the results to more general MDS codes and channels. Even when considering conventional MDS storage codes, or multiple-access compound wiretap channels where (12) is not achievable, connections still exist; this time, in the form of achievable upper and lower bounds with respect to repair overhead and S-DoF, respectively. The main point is that a storage code requiring a certain overhead for the repair of a node failure can still map to a channel where some S-DoF can be achieved and vice versa. The general performance bounds can be derived due to the fact that $\frac{\|\mathbf{s}\|_1}{T} \leq \|\mathbf{s}\|_\infty \leq \|\mathbf{s}\|_1 \leq T\|\mathbf{s}\|_\infty$, for any $\mathbf{s} \in \mathbb{R}^T$, which can be applied to the optimization sum-rank and max-rank cost functions.

*Lemma 3:* Let an $(n,k)_\beta$ MDS code $\mathbf{A}$ over $\mathbb{R}$ and $\delta_i$ be the overhead required to repair systematic node $i \in \{1,\ldots,k\}$. Then, $\mathbf{A}$ maps to an $(n-k,\beta)^{k-1}$ multiple-access compound wiretap channel $\mathbf{H} = \mathbf{P}_i\mathbf{A}$, where at least $[2-\delta_i]^+$ and at most $\frac{k-\delta_i}{k-1}$ S-DoF is achievable.

*Lemma 4:* Let an $(n,k)_\beta$ MDS code $\mathbf{A}$ over $\mathbb{R}$ and some set of repair matrices $\mathbf{R}_i$, $i \in \{1,\ldots,k\}$. Then, $\mathbf{A}$ maps to an $(n-k,\beta)^{k-1}$ multiple-access compound wiretap channel $\mathbf{H} = \mathbf{P}_i\mathbf{A}$, where $\eta\left((n-k,\beta)^{k-1},\mathbf{P}_i\mathbf{A},\mathbf{R}_i\right)$ S-DoF is achievable for $\mathbf{V} = \mathbf{R}_i$.

In the next example, we consider a code-to-channel mapping according to Lemma (4) and study the bounds derived by Lemma (3).

*Example 4*: Let an $(n,k)_\beta$ MDS code $\mathbf{A}$ and the repair overhead required for a failure of node $i \in \{1,\ldots,k\}$ is $\delta_i$ when using $\mathbf{R}_i$. Then, $\mathbf{A}$ maps to an $(n-k,\beta)^{k-1}$ channel $\mathbf{H} = \mathbf{P}_i\mathbf{A}$, $i \in \{1,\ldots,k\}$. For this channel we set $\mathbf{V} = \mathbf{R}_i$ and obtain

$$\eta\left((n-k,\beta)^{k-1},\mathbf{P}_i\mathbf{A},\mathbf{R}_i\right) = \eta\left((n-k,\beta)^{k-1},\mathbf{H},\mathbf{V}\right)$$
$$\geq \left[\frac{(n-k)\beta - \sum_{v=1}^{K-1}\text{rank}\left(\begin{bmatrix}\mathbf{H}_v^{(1)}\mathbf{V}^{(1)}\ldots\mathbf{H}_v^{(n-k)}\mathbf{V}^{(n-k)}\end{bmatrix}\right)}{(n-k)\beta}\right]^+$$
$$= [2-\delta_i]^+ = \begin{cases}\left[\frac{n-k-(k-1)}{n-k}\right]^+, & \delta_i = \frac{n-1}{n-k} \\ [2-k]^+, & \delta_i = k\end{cases}. \quad (14)$$

and

$$\eta\left((n-k,\beta)^{k-1},\mathbf{P}_i\mathbf{A},\mathbf{R}_i\right) = \eta\left((n-k,\beta)^{k-1},\mathbf{H},\mathbf{V}\right)$$
$$\leq \frac{(n-k)(k-1)\beta - \sum_{v=1}^{K-1}\text{rank}\left(\begin{bmatrix}\mathbf{H}_v^{(1)}\mathbf{V}^{(1)}\ldots\mathbf{H}_v^{(n-k)}\mathbf{V}^{(n-k)}\end{bmatrix}\right)}{(k-1)(n-k)\beta}$$
$$= \frac{(n-k)k\beta}{(k-1)(n-k)\beta} - \frac{\delta_i}{k-1}$$
$$= \frac{k-\delta_i}{k-1} = \begin{cases}1 - \frac{k+1}{(k-1)(n-k)}, & \delta_i = \frac{n-1}{n-k} \\ 0, & \delta_i = k\end{cases}. \quad (15)$$

For $k = 2$, (14) is tight for $\delta_i = \frac{n-1}{n-2}$. For any $k > 1$ and $\delta_i = k$, (15) is always tight. Here, we considered only the two extreme values of $\delta_i$.

We continue with performance bounds for channel-to-code mappings, for channels that do not necessarily achieve (12).

*Lemma 5:* Let an $(L,N)^{K-1}$ multiple-access compound wiretap channel $\mathbf{H}$, where $\eta$ S-DoF is achievable. Then, $\mathbf{H}$ maps to an $(L+K,K)_N$ MDS code $\mathbf{A} = \mathbf{H}$ over $\mathbb{R}$ and the overhead to repair node 1 is at most $1 + (K-1)(1-\eta)$ and at least $2 - \eta$.

*Lemma 6:* Let an $(L,N)^{K-1}$ multiple-access compound wiretap channel $\mathbf{H}$ and a set of beamforming matrices $\mathbf{V}$. Then, $\mathbf{H}$ maps to an $(L+K,K)_N$ MDS code $\mathbf{A} = \mathbf{H}$ over $\mathbb{R}$, where overhead $\delta\left((L+K,K)_N,\mathbf{H},\mathbf{V}\right)$ to repair node 1 is achievable for $\mathbf{R}_1 = \hat{\mathbf{V}}$.

We consider a code-to-channel mapping according to Lemma (6) and study the bounds derived by Lemma (5).

*Example 5*: Let an $(L,N)^{K-1}$ multiple-access compound wiretap channel $\mathbf{H}$, where $\eta$ S-DoF is achievable using $\mathbf{V}$. Then, $\mathbf{H}$ maps to an $(L+K,K)_N$ MDS code $\mathbf{A} = \mathbf{H}$. For

this code we set $\mathbf{R}_1 = \mathbf{V}$ and obtain

$$\delta\left((L+K,K)_N, \mathbf{H}, \mathbf{V}\right) = \delta\left((L+K,K)_N, \mathbf{A}, \mathbf{R}_1\right)$$
$$\leq \frac{LN + (K-1)\max_{u \in \{2,\ldots,K\}} \text{rank}\left(\left[\mathbf{A}_u^{(1)}\mathbf{R}_1^{(1)} \ldots \mathbf{A}_u^{(n-k)}\mathbf{R}_1^{(n-k)}\right]\right)}{LN}$$
$$= 1 + (K-1)(1-\eta) = \begin{cases} \frac{L+K-1}{L}, & \eta = \frac{L-1}{L} \\ K, & \eta = 0 \end{cases}. \quad (16)$$

and

$$\delta\left((L+K,K)_N, \mathbf{H}, \mathbf{V}\right) = \delta\left((L+K,K)_N, \mathbf{A}, \mathbf{R}_1\right)$$
$$\geq \frac{LN + \max_{u \in \{2,\ldots,K\}} \text{rank}\left(\left[\mathbf{A}_u^{(1)}\mathbf{R}_1^{(1)} \ldots \mathbf{A}_u^{(n-k)}\mathbf{R}_1^{(n-k)}\right]\right)}{LN}$$
$$= 2 - \eta = \begin{cases} 1 + \frac{1}{L}, & \eta = \frac{L-1}{L} \\ 2, & \eta = 0 \end{cases}. \quad (17)$$

Observe that (16) is tight for both cases and all any code considered, but (17) not. Again, we considered the two extreme values of $\eta$.

We need to note that by the previous lemmas, it is implicit that having a black-box solving either $\mathcal{R}$ or $\mathcal{V}$ is useful for both the repair minimization and S-DoF maximization problem, since any solution to one problem plugs in as a solution to the corresponding mapped one in the analogous setting and certain approximations are guaranteed.

To conclude, in this section we established connections between the repair overhead minimization of MDS storage codes and the S-DoF maximization of multiple-access compound wiretap channels.

## V. CONCLUSIONS

In this work we showed that the repair overhead minimization for optimal MDS storage codes is equivalent to the S-DoF maximization in a multiple-access compound wiretap channel. The reverse holds for multiple-access compound wiretap channels where (12) is tight. The framework for the connections was established by restating the two problems as rank constrained sum-rank and max-rank minimizations, respectively. Through this framework we established code-to-channel and channel-to-code mappings. Using such a mapping, we determined the maximum S-DoF in the single antenna, multiple-access compound wiretap channel. We also extended our results to conventional MDS storage codes and more general channels.

## APPENDIX

**Proof of (11)**: Let

$$\mathbf{S} \triangleq \frac{1}{\sqrt{P}}\left[\mathbf{H}^{(1)}\mathbf{V}^{(1)}\ldots\mathbf{H}^{(L)}\mathbf{V}^{(L)}\right] \in \mathbb{R}^{LN \times LN},$$
$$\mathbf{S}_{ev} \triangleq \frac{1}{\sqrt{P}}\left[\mathbf{H}_{ev}^{(1)}\mathbf{V}^{(1)}\ldots\mathbf{H}_{ev}^{(L)}\mathbf{V}^{(L)}\right] \in \mathbb{R}^{LN \times LN}$$

for all $v \in \{1,\ldots,K-1\}$. Then,

$$L \cdot N \cdot \eta\left((L,N)^{K-1}, \mathbf{H}, \mathbf{V}\right)$$
$$= \lim_{P \to \infty} \frac{\frac{1}{2}\log \det\left(\mathbf{I}_{M_r} + \frac{1}{\sigma^2}\sum_{l=1}^{L}\mathbf{H}^{(l)}\mathbf{V}^{(l)}\left(\mathbf{H}^{(l)}\right)^T\left(\mathbf{V}^{(l)}\right)^T\right)}{\frac{1}{2}\log\left(\frac{P}{\sigma^2}\right)}$$
$$= \lim_{P \to \infty} \frac{\frac{1}{2}\log\det\left(\frac{P}{\sigma^2}\mathbf{S}\mathbf{S}^T\right) - \max_{v \in \{1,\ldots,K-1\}}\frac{1}{2}\log\det\left(\frac{P}{\sigma^2}\mathbf{S}_{ev}\mathbf{S}_{ev}^T\right)}{\frac{1}{2}\log\left(\frac{P}{\sigma^2}\right)}$$
$$= \lim_{P \to \infty} \frac{\sum_{i=1}^{\text{rank}(\mathbf{S})}\frac{1}{2}\log\left(\frac{P}{\sigma^2}\sigma_i^S\right) - \max_{v \in \{1,\ldots,K-1\}}\sum_{i=1}^{\text{rank}(\mathbf{S}_{ev})}\frac{1}{2}\log\left(\frac{P}{\sigma^2}\sigma_i^v\right)}{\frac{1}{2}\log\left(\frac{P}{\sigma^2}\right)}$$
$$= \lim_{P \to \infty} \frac{\text{rank}(\mathbf{S})\frac{1}{2}\log\left(\frac{P}{\sigma^2}\right) - \max_{v \in \{1,\ldots,K-1\}}\text{rank}(\mathbf{S}_{ev})\frac{1}{2}\log\left(\frac{P}{\sigma^2}\right)}{\frac{1}{2}\log\left(\frac{P}{\sigma^2}\right)}$$
$$= \text{rank}\left(\left[\mathbf{H}^{(1)}\mathbf{V}^{(1)}\ldots\mathbf{H}^{(L)}\mathbf{V}^{(L)}\right]\right) - \max_{v \in \{1,\ldots,K-1\}}\text{rank}\left(\left[\mathbf{H}_{ev}^{(1)}\mathbf{V}^{(1)}\ldots\mathbf{H}_{ev}^{(L)}\mathbf{V}^{(L)}\right]\right).$$

The previous example is possible when the maximum singular values of the channel matrices are not scaling with the power $P$. Namely, for the $i$th largest singular of matrix $\mathbf{S}$, $\sigma_i$ we obtain

$$\sigma_i \triangleq \frac{1}{P}\max_{\|\mathbf{x}\|_2 = 1}\left\|\left[\mathbf{H}^{(1)}\mathbf{V}^{(1)}\ldots\mathbf{H}^{(L)}\mathbf{V}^{(L)}\right]\mathbf{x}\right\|_2^2$$
$$= \frac{1}{P}\max_{\|\mathbf{x}\|_2 = 1}\left\|\sum_{l=1}^{L}\mathbf{H}^{(l)}\mathbf{V}^{(l)}\mathbf{x}^{(l)}\right\|_2^2$$
$$\leq \frac{1}{P}\sum_{l=1}^{L}\max_{\|\mathbf{V}^{(l)}\mathbf{x}^{(l)}\|_2 \leq \sqrt{P}}\left\|\mathbf{H}^{(l)}\mathbf{x}_l\right\|_2^2 \leq \frac{1}{P}\sum_{l=1}^{L}P\max_{\|\mathbf{x}_l\|_2 = 1}\left\|\mathbf{H}^{(l)}\mathbf{x}_l\right\|_2^2$$
$$= L\max_{i \in \{1,\ldots,LN\}, l \in \{1,\ldots,L\}}\sigma_i\left(\mathbf{H}^{(l)}\right)$$

for $\mathbf{x} = \left[\left(\mathbf{x}^{(1)}\right)^T \ldots \left(\mathbf{x}^{(L)}\right)^T\right]^T$, $\mathbf{x}_l \in \mathbb{R}^{LN \times 1}$, for $l \in \{1,\ldots,L\}$. In the same fashion, we have that the $i$th largest singular value of matrix $\mathbf{S}_{ev}$ is upper bounded. Hence, the singular values are not scaling with $P$ as long as the largest singular values of the channels are not scaling with $P$. □

**Proof of Theorem 1**: Let an optimal $(n,k)_\beta$ MDS code $\mathbf{A}$ and a systematic node $i \in \{1,\ldots,k\}$ fail. Then, setting $L = n-k$, $N = \beta$, $K = k$, and $\mathbf{H} = \mathbf{P}_i\mathbf{A}$, accounts for constructing an $(n-k,\beta)^{k-1}$ multiple-access compound wiretap channel instance $\mathbf{H}$. Then, the following mappings hold

$$\left[\mathbf{H}^{(1)}\mathbf{V}^{(1)}\ldots\mathbf{H}^{(n-k)}\mathbf{V}^{(n-k)}\right] = \left[\mathbf{A}_i^{(1)}\mathbf{V}^{(1)}\ldots\mathbf{A}_i^{(n-k)}\mathbf{V}^{(n-k)}\right]$$
$$= \left[\mathbf{A}_i^{(1)}\mathbf{R}_i^{(1)}\ldots\mathbf{A}_i^{(n-k)}\mathbf{R}_i^{(n-k)}\right]$$

and

$$\left[\mathbf{H}_{ev}^{(1)}\mathbf{V}^{(1)}\ldots\mathbf{H}_{ev}^{(n-k)}\mathbf{V}^{(n-k)}\right] = \left[\mathbf{A}_u^{(1)}\mathbf{V}^{(1)}\ldots\mathbf{A}_u^{(n-k)}\mathbf{V}^{(n-k)}\right]$$
$$= \left[\mathbf{A}_u^{(1)}\mathbf{R}_i^{(1)}\ldots\mathbf{A}_u^{(n-k)}\mathbf{R}_i^{(n-k)}\right]$$

for any $\mathbf{V} = \mathbf{R}_i$ and $v = \begin{cases} u, & u \in \{1,\ldots,i-1\}, \\ u-1, & u \in \{i+1,\ldots,k\}. \end{cases}$
Hence, the full rank constraint of $\mathcal{R}\left((n,k)_\beta, \mathbf{P}_i\mathbf{A}\right)$ is equivalent to the full rank constraint of $\mathcal{V}\left((n-k,\beta)^{k-1}, \mathbf{P}_i\mathbf{A}\right)$.

The objectives are also equivalent since

$$\min_{\mathbf{R}_i} \sum_{u=1,u\neq i}^{k} \operatorname{rank}\left(\left[\mathbf{A}_u^{(1)}\mathbf{R}_i^{(1)} \ldots \mathbf{A}_u^{(n-k)}\mathbf{R}_i^{(n-k)}\right]\right)$$
$$= (k-1)\min_{\mathbf{R}_i} \max_{u\in\{1,\ldots,k\}\setminus i} \operatorname{rank}\left(\left[\mathbf{A}_u^{(1)}\mathbf{R}^{(1)} \ldots \mathbf{A}_u^{(n-k)}\mathbf{R}_i^{(n-k)}\right]\right)$$
$$= (k-1)\min_{\mathbf{V}} \max_{v\in\{1,\ldots,k-1\}} \operatorname{rank}\left(\left[\mathbf{H}_{ev}^{(1)}\mathbf{V}^{(1)} \ldots \mathbf{H}_{ev}^{(n-k)}\mathbf{V}^{(n-k)}\right]\right)$$
$$= (k-1)\beta.$$

for optimal MDS codes. $\square$

**Proof of Theorem 2**: Let an $(L,N)^{K-1}$ multiple-access compound wiretap channel $\mathbf{H}$, where $\frac{L-1}{L}$ S-DoF is achievable. Then, setting $n = L+K$, $K = k$, $\beta = N$, and $\mathbf{A} = \mathbf{H}$, accounts for constructing an $(L+K,K)_N$ MDS code $\mathbf{A}$. Then, the following mappings hold

$$\left[\mathbf{A}_1^{(1)}\mathbf{R}_1^{(1)} \ldots \mathbf{A}_1^{(L)}\mathbf{R}_1^{(L)}\right] = \left[\mathbf{H}^{(1)}\mathbf{R}_1^{(1)} \ldots \mathbf{H}^{(L)}\mathbf{R}_1^{(L)}\right]$$
$$= \left[\mathbf{H}^{(1)}\mathbf{V}^{(1)} \ldots \mathbf{H}^{(L)}\mathbf{V}^{(L)}\right]$$

and

$$\left[\mathbf{A}_u^{(1)}\mathbf{R}_i^{(1)} \ldots \mathbf{A}_u^{(L)}\mathbf{R}_i^{(n-k)}\right] = \left[\mathbf{H}_{ev}^{(1)}\mathbf{R}_1^{(1)} \ldots \mathbf{H}_{ev}^{(L)}\mathbf{R}_1^{(L)}\right]$$
$$= \left[\mathbf{H}_{ev}^{(1)}\mathbf{V}^{(1)} \ldots \mathbf{H}_{ev}^{(L)}\mathbf{V}^{(L)}\right]$$

for any $\mathbf{V} = \mathbf{R}_i$ and $v = u-1$. The full rank constraint of $\mathcal{V}\left((n-k,\beta)^{k-1}, \mathbf{P}_i\mathbf{A}\right)$ is equivalent to the full rank constraint of $\mathcal{R}\left((n,k)_\beta, \mathbf{A}\right)$. The objectives are also equivalent since

$$(K-1)\min_{\mathbf{V}} \max_{v\in\{1,\ldots,K-1\}} \operatorname{rank}\left(\left[\mathbf{H}_{ev}^{(1)}\mathbf{V}^{(1)} \ldots \mathbf{H}_{ev}^{(L)}\mathbf{V}^{(n-k)}\right]\right)$$
$$= (K-1)\min_{\mathbf{R}_1} \max_{u\in\{2,\ldots,k\}} \operatorname{rank}\left(\left[\mathbf{A}_u^{(1)}\mathbf{R}_1^{(1)} \ldots \mathbf{A}_u^{(L)}\mathbf{R}_1^{(L)}\right]\right)$$
$$= \min_{\mathbf{R}_1} \sum_{u=2}^{k} \operatorname{rank}\left(\left[\mathbf{A}_u^{(1)}\mathbf{R}_1^{(1)} \ldots \mathbf{A}_u^{(L)}\mathbf{R}_1^{(n-k)}\right]\right)$$
$$= (k-1)\beta.$$

for channels where $\frac{L-1}{L}$ total S-DoF is achievable. $\square$

**Proof of lemma 3**: Let an $(n,k)_\beta$ MDS code $\mathbf{A}$ over $\mathbb{R}$ with minimum overhead to repair systematic node $i$ equal to $\delta_i$ achieved by $\mathbf{R}_i$. Then, the maximum S-DoF of the $(n-k,\beta)^{k-1}$ channel $\mathbf{H} = \mathbf{P}_i\mathbf{A}$ is upper bounded by

$$\eta\left((n-k,\beta)^{k-1}, \mathbf{P}_i\mathbf{A}, \mathbf{R}_i\right) = \eta\left((n-k,\beta)^{k-1}, \mathbf{H}, \mathbf{V}\right)$$
$$= \frac{(n-k)\beta - \max_{v\in\{1,\ldots,k-1\}} \operatorname{rank}\left(\left[\mathbf{H}_v^{(1)}\mathbf{V}^{(1)} \ldots \mathbf{H}_v^{(n-k)}\mathbf{V}^{(n-k)}\right]\right)}{(n-k)\beta}$$
$$\geq \left[\frac{(n-k)\beta - \sum_{v=1}^{k-1} \operatorname{rank}\left(\left[\mathbf{H}_v^{(1)}\mathbf{V}^{(1)} \ldots \mathbf{H}_v^{(n-k)}\mathbf{V}^{(n-k)}\right]\right)}{(n-k)\beta}\right]^+$$
$$= [2-\delta_i]^+$$

and lower bounded by

$$\eta\left((n-k,\beta)^{k-1}, \mathbf{P}_i\mathbf{A}, \mathbf{R}_i\right) = \eta\left((n-k,\beta)^{k-1}, \mathbf{H}, \mathbf{V}\right)$$
$$= \frac{(n-k)\beta - \max_{v\in\{1,\ldots,k-1\}} \operatorname{rank}\left(\left[\mathbf{H}_v^{(1)}\mathbf{V}^{(1)} \ldots \mathbf{H}_v^{(n-k)}\mathbf{V}^{(n-k)}\right]\right)}{(n-k)\beta}$$
$$\leq \frac{(k-1)(n-k)\beta - \sum_{v=1}^{k-1} \operatorname{rank}\left(\left[\mathbf{H}_v^{(1)}\mathbf{V}^{(1)} \ldots \mathbf{H}_v^{(n-k)}\mathbf{V}^{(n-k)}\right]\right)}{(k-1)(n-k)\beta}$$
$$= \frac{(n-k)k\beta}{(k-1)(n-k)\beta} - \frac{\delta_i}{k-1}$$
$$= \frac{k}{k-1} - \frac{\delta_i}{k-1}.$$

for $\mathbf{V} = \mathbf{R}_i$, $i \in \{1, \ldots, k\}$. $\square$

**Proof of lemma 4**: Let an $(n,k)_\beta$ MDS code $\mathbf{A}$ over $\mathbb{R}$, and the set $\mathbf{R}_i$ of repair matrices for the repair of systematic node $i \in \{1, \ldots, k\}$, satisfying the constraint of $\mathcal{R}_i((n,k)_\beta, \mathbf{P}_i\mathbf{A})$. We define the $(n-k,\beta)^{k-1}$ channel instance $\mathbf{H} = \mathbf{P}_i\mathbf{A}$. Then, by setting $\mathbf{V} = \mathbf{R}_i$ we achieve the following S-DoF for $\mathbf{H}$

$$\frac{\operatorname{rank}\left(\left[\mathbf{A}_u^{(1)}\mathbf{R}_i^{(1)} \ldots \mathbf{A}_u^{(n-k)}\mathbf{R}_i^{(n-k)}\right]\right)}{(n-k)\beta}$$
$$- \frac{\max_{u\in\{1,\ldots,k\}\setminus i} \operatorname{rank}\left(\left[\mathbf{A}_u^{(1)}\mathbf{R}_i^{(1)} \ldots \mathbf{A}_u^{(n-k)}\mathbf{R}_i^{(n-k)}\right]\right)}{(n-k)\beta}$$
$$\frac{(n-k)\beta - \max_{u\in\{1,\ldots,k\}\setminus i} \operatorname{rank}\left(\left[\mathbf{A}_u^{(1)}\mathbf{R}_i^{(1)} \ldots \mathbf{A}_u^{(n-k)}\mathbf{R}_i^{(n-k)}\right]\right)}{(n-k)\beta}$$
$$= \eta\left((n-k,\beta)^{k-1}, \mathbf{P}_i\mathbf{A}, \mathbf{R}_i\right) = \eta\left((n-k,\beta)^{k-1}, \mathbf{H}, \mathbf{V}\right).$$

$\square$

**Proof of lemma 5**: Let an $(L,N)^{K-1}$ multiple-access compound wiretap channel $\mathbf{H}$ where $\eta$ S-DoF is achievable by $\mathbf{V}$. Then, let us define the $(L+K,K)_N$ MDS code $\mathbf{A} = \mathbf{H}$. For this code we have

$$\delta((L+K,K)_N, \mathbf{H}, \mathbf{V}) = \delta((L+K,K)_N, \mathbf{A}, \mathbf{R}_1)$$
$$\leq \frac{LN + (K-1)\max_{u\in\{2,\ldots,K\}} \operatorname{rank}\left(\left[\mathbf{A}_u^{(1)}\mathbf{R}_1^{(1)} \ldots \mathbf{A}_u^{(L)}\mathbf{R}_1^{(L)}\right]\right)}{LN}$$
$$= 1 + (K-1)(1-\eta)$$

and

$$\delta((L+K,K)_N, \mathbf{H}, \mathbf{V}) = \delta((L+K,K)_N, \mathbf{A}, \mathbf{R}_1)$$
$$\geq \frac{LN + \max_{u\in\{2,\ldots,K\}} \operatorname{rank}\left(\left[\mathbf{A}_u^{(1)}\mathbf{R}_1^{(1)} \ldots \mathbf{A}_u^{(L)}\mathbf{R}_1^{(L)}\right]\right)}{LN} = 2 - \eta.$$

for $\mathbf{R}_1 = \mathbf{V}$. $\square$

**Proof of lemma 6**: Let an $(L,N)^{K-1}$ multiple-access compound wiretap channel $\mathbf{H}$ and the set $\mathbf{V}$ of beamforming matrices, satisfying the constraint of $\mathcal{V}_i((L,N)^{K-1}, \mathbf{H})$. We define the $(L+K,K)_N$ MDS code instance $\mathbf{A} = \mathbf{H}$. Then, by setting $\mathbf{R}_1 = \mathbf{V}$ we achieve the following overhead for the

repair of node 1

$$\frac{\text{rank}\left(\left[\mathbf{H}^{(1)}\mathbf{V}^{(1)}\ldots\mathbf{H}^{(L)}\mathbf{V}^{(L)}\right]\right)}{LN}$$
$$+\frac{\sum_{v=1}^{K-1}\text{rank}\left(\left[\mathbf{H}_v^{(1)}\mathbf{V}^{(1)}\ldots\mathbf{H}_v^{(L)}\mathbf{V}_i^{(L)}\right]\right)}{LN}$$
$$=\frac{LN+\sum_{v=1}^{K-1}\text{rank}\left(\left[\mathbf{H}_v^{(1)}\mathbf{V}^{(1)}\ldots\mathbf{H}_v^{(L)}\mathbf{V}^{(L)}\right]\right)}{LN}$$
$$=\delta\left((n,k)_\beta,\mathbf{H},\mathbf{V}\right).$$

□

### REFERENCES

[1] M. Maddah-Ali, A. Motahari, and A. Khandani, "Communication over MIMO X channels: Interference alignment, decomposition, and performance analysis," in *IEEE Trans. on Inform. Theory*, pp. 3457–3470, Aug. 2008.

[2] V. R. Cadambe and S. A. Jafar, "Interference alignment and degrees of freedom of the $K$-user interference channel," in *IEEE Trans. on Inform. Theory*, vol. 54, pp. 3425–3441, Aug. 2008.

[3] S. Jafar and S. Shamai, "Degrees of freedom region for the mimo X channel," in *IEEE Trans. on Inform. Theory*, vol. 54, Jan. 2008.

[4] T. Gou and S. Jafar, "Degrees of freedom of the $K$ user mimo interference channel," in *42nd Asilomar Conf. on Signals, Systems and Computers 2008*, (Pacific Grove, CA, USA), Oct. 2008.

[5] C. Suh and D. Tse, "Interference alignment for cellular networks," in *Proceedings of 40th Annual Allerton Conf. on Commun., Control, and Computing 2008*, (Urbana-Champaign, IL, USA), Sep. 2008.

[6] O. Koyluoglu, H. E. Gamal, L. Lai, and H. V. Poor, "On the secure degrees of freedom in the k-user gaussian interference channel," May 2008. Available online at arXiv:0805.1340v1 [cs.IT].

[7] A. Khisti, "Interference alignment for the multi-antenna compound wiretap channel," Mar. 2010. Available online at arXiv:1002.4548v2 [cs.IT].

[8] G. Bagherikaram, A. S. Motahari, and A. K. Khandani, "On the secure degrees-of-freedom of the multiple-access-channel," Mar. 2010. Available online at arXiv:1003.0729v1 [cs.IT].

[9] M. Kobayashi, P. Piantanida, S. Yang, and S. Shamai, "On the secrecy degress of freedom of the multi-antenna block fading wiretap channels," May 2010. Available online at arXiv:1001.2410v2 [cs.IT].

[10] X. He and A. Yener, "The gaussian many-to-1 interference channel with confidential messages," May 2010. Available online at arXiv:1005.0624v1 [cs.IT].

[11] R. Bassily and S. Ulukus, "Ergodic secret alignment for the fading multiple access wiretap channel," in *Proc. ICC 2010*, (Cape Town, South Africa), May 2010.

[12] A. G. Dimakis, P. B. Godfrey, Y. Wu, M. J. Wainwright, and K. Ramchandran, "Network coding for distributed storage systems," in *IEEE Transactions on Information Theory* (to appear).

[13] Y. Wu and A. G. Dimakis, "Reducing repair traffic for erasure coding-based storage via interference alignment," in *IEEE Int. Symp. on Information Theory (ISIT)*, (Seoul, Korea), July 2009.

[14] A. G. Dimakis, K. Ramchandran, Y. Wu, and C. Suh, "A survey on network codes for distributed storage," Available online at arXiv:1004.4438v1 [cs.IT].

[15] D. Cullina, A. G. Dimakis, and T. Ho, "Searching for minimum storage regenerating codes," Oct. 2009. Available online arXiv:0910.2245v1 [cs.IT].

[16] K. V. Rashmi, N. B. Shah, P. V. Kumar, and K. Ramchandran, "Explicit construction of optimal exact regenerating codes for distributed storage," Oct. 2009. Available online at arXiv:0906.4913v2 [cs.IT].

[17] K. V. Rashmi, N. B. Shah, P. V. Kumar, and K. Ramchandran, "Interference alignment as a tool in network coding as applied to distributed storage," in *Proc. National Conf. on Commun.*, (Chennai, India), Jan. 2010.

[18] N. B. Shah, K. V. Rashmi, P. V. Kumar, and K. Ramchandran, "The miser code: an mds distributed storage code that minimizes repair bandwidth for systematic nodes through interference alignment," May 2010. Available online at arXiv:0908.2984v2 [cs.IT].

[19] V. R. Cadambe, S. A. Jafar, and H. Maleki, "Distributed data storage with minimum storage regenerating codes - exact and functional repair are asymptotically equally efficient," Apr. 2010. Available online at arXiv:1004.4299v1 [cs.IT].

[20] C. Suh and K. Ramchandran, "On the existence of optimal exact-repair mds codes for distributed storage," Apr. 2010. Available online at arXiv:1004.4663v1 [cs.IT].

[21] D. Papailiopoulos and A. Dimakis, "Interference alignment as a rank constrained rank minimization," in *IEEE GLOBECOM 2010*, (Miami, FL, USA).